\documentstyle[12pt]{article}

\newcommand{\be}{\begin{eqnarray}}
\newcommand{\ee}{\end{eqnarray}}
\newcommand{\e}{\varepsilon}
\newcommand{\D}{\partial}

\makeatletter
\@addtoreset{equation}{section}
\makeatother

\begin{document}
\gdef\journal#1, #2, #3, 1#4#5#6{{#1~}{#2} (1#4#5#6) #3}
\gdef\ibid#1, #2, 1#3#4#5{{#1} (1#3#4#5) #2}

\begin{flushright}
OSLO-TP 2-97\\
cond-mat/9701154
\end{flushright}

\vspace{1cm}

{\centerline{\Large\bf Statistical mechanics and thermodynamics}}
\smallskip
{\centerline {\Large\bf for multispecies exclusion statistics }}
\vskip 1cm
\centerline {\large Serguei B. Isakov$^{a,b}$,
Stefan Mashkevich$^{a,c}$}
\vskip .5cm
\centerline{ \em $^a$Senter for H\o yere Studier,
Drammensveien 78, 0271 Oslo, Norway}
\smallskip
\centerline {\it $^b$Department  of Physics,  University of Oslo}
\centerline {\it P.O. Box 1048 Blindern, 0316 Oslo,
Norway\footnote{Present address. Email: serguei.isakov@fys.uio.no}}
\smallskip
\centerline {\it $^c$Institute for Theoretical Physics, 252143
Kiev, Ukraine\footnote{Present address. Email: mash@phys.ntnu.no}}

\vskip 1cm

\centerline {\bf Abstract}

\vskip .5cm

Statistical mechanics and thermodynamics for ideal fractional
exclusion stati\-stics with mutual statistical interactions
is studied systematically.
We discuss properties of the single-state partition functions
and derive the general form of the cluster expansion.
Assuming a certain scaling of the single-particle partition
functions, relevant to systems of noninteracting particles
with various dispersion laws, both in a box and
in an external harmonic potential, we derive a
unified form of the virial expansion. For the case of
a symmetric statistics matrix at a constant density of states,
the thermodynamics is analyzed completely.
We solve the microscopic problem of multispecies anyons
in the lowest Landau level for arbitrary values
of particle charges and masses (but the same sign of charges).
Based on this, we derive the equation of state which has the form
implied by exclusion statistics, with the statistics matrix
coinciding with the exchange statistics matrix of anyons.
Relation to one-dimensional integrable models is discussed.

\noindent


\newpage

\section{Introduction}
In recent years it has been appreciated that there exists a nontrivial
possibility of mutual quantum statistics,
or ``statistical interaction'' between distinguishable particles.
One example is multispecies exclusion statistics \cite{H91},
which postulates that inserting into the system
a particle of species $b$ reduces the number $D_a$ of single-particle
states available for species $a$ by some $g_{ab}$. The quantities
$g_{ab}$ constitute the exclusion statistics matrix
(later referred to as the statistics matrix)
and are called mutual statistics parameters for $a\neq b$.

Another example is multispecies  exchange statistics, or
multispecies anyons \cite{BFHI91,W92,SSS}.
Their definition is the following: (i) when two particles $a$ are
interchanged, the two-body wave function acquires a phase factor
$\exp[{\rm i}\pi\alpha_{aa}]$; (ii) when particle $a$ encircles
particle $b$, it acquires a phase factor
$\exp[2{\rm i}\pi\alpha_{ab}]$.
The matrix $\{\alpha_{ab}\}$ will be called exchange
statistics matrix.

In fact, there is a connection between
the two models, which was first established for a single species:
The thermodynamic quantities for a system of anyons (in a box)
confined to the lowest Landau level (LLL) of a strong external
magnetic field \cite{dVO94} are the same as those of
exclusion statistics particles, all having the same
energy (energy of the LLL) upon identification
$g=\alpha$ \cite{Wu94}. This correspondence was extended to the
multispecies model
in the particular case when the electric charges of all the
species are equal; one then has
$g_{ab} = \alpha_{ab}$ \cite{SSS}.

The relation between the two models was also
discussed in the context of an algebraic approach to statistics
(aimed at searching for an algebra of observables for
identical particles). The algebraic definition of fractional
statistics in one dimension \cite{LM-HeisQuant}
applies to anyons in the LLL, where the dynamics
is effectively one-dimensional \cite{HLM}.
On the other hand, this definition suggests a realization
of 1D fractional statistics by systems
with inverse square interactions \cite{LM-HeisQuant}
(see also \cite{Poly-NPB89}), which has been justified
by interpreting the thermodynamics of the
Calogero and Sutherland models \cite{Calo, Sutherland}
as that of systems of noninteracting particles \cite{I-IJMPA}.
The statistical distribution for 1D fractional statistics
found in this way \cite{I-IJMPA} turns out to be the same
as that derived from Haldane's multiplicity formula
for exclusion statistics \cite{I-MPLB,Wu94}.

In addition to the above, (single-species) integrable models of the
Calogero-Sutherland type were directly related to
exclusion statistics  \cite{BWu,MS94,Ha}.
It is therefore natural to expect
that there should exist multispecies
generalizations of those models that would correspond to
multispecies exclusion statistics.

It is generally believed that excitations in the
fractional quantum Hall effect (FQHE) can be described
either as anyons (by a Berry phase argument \cite{ASW84})
or as exclusion statistics particles (by a state-counting
argument \cite{H91,ES-Hall}).
In particular, when excitations with charges of different
magnitudes are present --- in multilayer systems \cite{W92}
and in the hierarchical structure \cite{hier} --- the
multispecies model is applicable.
Another problem of interest is that of FQHE quasielectrons
and quasiholes, again viewed as two species.
By counting the states, their off-diagonal exclusion
statistics parameters were recently shown \cite{offdiag}
to be antisymmetric; on the other
hand, the exchange statistics matrix is  always
symmetric by definition. How the two descriptions match
in this case, is not clear by now.

Therefore, developing statistical mechanics for mutual statistics is of
interest both theoretically and phenomenologically.
Thus far, some progress has been achieved on the issue
\cite{I-MPLB,dVO95,SSS}, generalizing the results of the
better-understood single-species case,
where the thermodynamics was studied for a system in a
box \cite{Wu94,MS94,NW94,IAMP} as well as in a harmonic potential
\cite{MS94,BhSen,IsO}.

This paper is devoted to systematically studying statistical
mechanics of particles of arbitrary number of species obeying
exclusion statistics and calculat\-ing their thermodynamic
quantities.

In Sec.~\ref{PARTFUN}
we introduce the single-state grand canonical partition function
and prove that it factorizes into a product of
partition functions corresponding to separate species.
We infer a general formula
for the coefficients of the expansion of its associated single-state
thermodynamic potential in powers of the single-state Gibbs factors.
Those coefficients, multiplied by dimensional factors depending on the
dispersion law of the particles and the dimension of space, give the
cluster coefficients, which are discussed in Sec.~\ref{CLVIEXP}.
Assuming a certain temperature scaling of the single-particle
partition functions --- which holds for systems of noninteracting
particles with various dispersion laws, both in a box
and in an external harmonic potential, --- we also derive a
unified form of the virial expansion, calculating explicitly a few
first coefficients.

Section \ref{SYMMAT} is devoted to a specific case
when the statistics matrix is symmetric and in addition
the density of single-particle states is constant in energy.
This case can be analyzed completely, yielding a closed form
of the equation of state, which may be expanded to give both
low-temperature and low-density series.
In Sec.~\ref{EXAMP}
we discuss various physical examples of exclusion statistics for
systems of noninteracting particles both in a box and in an external
harmonic potential.

In Sec.~\ref{LLL} we address the problem of
multispecies anyons in the LLL at different values
(but the same sign) of charges and different masses
(hence different cyclotron frequencies)
for different species. In Subsec.~\ref{SPECTRUM} we solve
this problem exactly, exposing anyons of different species
to harmonic potentials with different frequencies
such that the resulting level spacing is the same for all the species.
In Subsec.~\ref{EQNOFSTATE}
we derive the equations of state for multispecies anyons
in the LLL both in a harmonic potential and in a box.
It has the form implied by exclusion statistics,
with the statistics matrix coinciding with
the exchange statistics matrix of anyons.
Relation to exclusion statistics is also demonstrated
by a semiclassical argument in Subsec.~\ref{SEMICL}.

We conclude with remarks on searching for microscopic integrable
models for multispecies exclusion statistics.

\section{Partition functions}
\label{PARTFUN}

The definition of exclusion statistics is the following \cite{H91}.
If $G_a$ is the initial single-particle Hilbert space dimension
for species $a$ and $N_a$ the number of particles of that species,
then (i) the reduced Hilbert space dimension is
\begin{equation}
D_a=G_a-\sum_b g_{ab}(N_b-\delta_{ab}) \;,
\label{dimH}
\end{equation}
and (ii) the many-particle statistical weights (multiplicities)
are determined as
\be
W=\prod_a \frac{(D_a+N_a-1)!}{N_a!(D_a-1)!}\;.
\label{StatWeight}\ee
Applying (\ref{dimH})--(\ref{StatWeight}) locally in phase space, so
that the total multiplicity is
\be
W=\prod_i W^{(i)} \; ,
\label{local}\ee
where $i$ labels groups of single-particle states of nearby energy
$\e^{(i)}$,
and $ W^{(i)}$ is given by
(\ref{StatWeight}) with the changes $N_a\to N_a^{(i)}$,
$G_a\to G_a^{(i)}$, and $D_a\to D_a^{(i)}$,
makes the definition of {\em ideal} fractional
exclusion statistics for many species \cite{I-MPLB,Wu94}
(see also a related discussion of systems with internal degrees of
freedom \cite{FK-PRB}).

The above equations make the starting point for constructing
statistical mechanics.
Assuming that $G_a^{(i)}$ does not depend on $a$,\footnote{
This holds for a wide class of practically interesting systems
(see below, Sec.~\ref{EXAMP}).
However, the case of $G_a^{(i)}$ being $a$ dependent
could be worked out as well; see, e.g.,
Eqs.~(\ref{clusfin})--(\ref{eqst-LLL-box})
and the discussion thereafter.}
one deduces equations for the average occupation numbers
$n_a^{(i)}=N_a^{(i)}/G_a^{(i)}$
in the thermo\-dynam\-ic limit, which can be conveniently written
as \cite{I-MPLB,Wu94}
\be
\frac{w_a^{(i)}}{ \prod_b (1-w_b^{(i)})^{g_{ba}} }=x_a^{(i)}\;,
\label{basic}\ee
where $x_a^{(i)}=e^{\beta (\mu_a-\e_a^{(i)})}$ and
\be
w_a^{(i)}=\frac{n_a^{(i)}}{1+n_a^{(i)}-\sum_b g_{ab}n_b^{(i)}}\;.
\label{w-n}\ee
It is natural to introduce the single-state
grand canonical partition function $\xi^{(i)}$,
which determines the occupation numbers as \cite{I-MPLB}
\be
n_a^{(i)}=x_a^{(i)} \frac{\D}{\D x_a^{(i)}}\ln \xi^{(i)} \;.
\label{n-xi}\ee

It was observed \cite{SSS} that
for a symmetric statistics matrix, $g_{ab}=g_{ba}$,
the partition function $\xi^{(i)}$ factorizes as
\be
\xi^{(i)}=\prod_a \xi_a^{(i)} \;,
\label{xi-factor}\ee
where the grand partition function $\xi_a^{(i)}$ for particles of
species $a$ is connected with the occupation numbers by
[see (\ref{w-n})]
\be
\xi_a^{(i)}=\frac{1}{1-w_a^{(i)}} \;.
\label{xi-w}\ee

The relations (\ref{w-n})--(\ref{xi-w}) were found in Ref.~\cite{SSS}
by explicitly expanding (for the case of two species) the
functions $\xi_a^{(i)}$, $\xi^{(i)}$, and $n_a^{(i)}$ in powers of the
Gibbs factors
$\{x_a^{(i)}\}$. We observe here that the above relations
can be put on a more general ground. Namely,
in the generic case of a nonsymmetric $g_{ab}$,
from Eq.~(\ref{basic}) we find
[we drop the superscript $(i)$ from now on]
\be
\frac{\D w_a}{\D x_b}=\frac{1}{x_b}( D^{-1})_{ab} \;,
\label{dw/dx}\ee
where
\be
D_{ab}=\frac{\delta_{ab}}{w_a}+\frac{g_{ba}}{1-w_{b}} \;.
\label{D}\ee
Equations (\ref{n-xi})--(\ref{xi-w}) then imply that
$D_{ab}$ relates the partition functions to the occupation numbers:
\be
\sum_a n_a D_{ab}=\xi_b \:.
\label{D-rel}\ee
Substituting (\ref{xi-w}) and (\ref{D}) herein recovers (\ref{w-n}).
This shows that (\ref{n-xi})--(\ref{xi-w}) are
consistent with (\ref{basic}), (\ref{w-n}) in the generic case
(any dispersion law and a nonsymmetric matrix $g_{ab}$).

Equations (\ref{basic}), (\ref{xi-factor}), and (\ref{xi-w})
implicitly determine $\xi$ as a function of the Gibbs factors
$\{ x_a \}$. In particular, expanding
\be
\ln \xi= \sum_{k_1 \ldots \,k_s}f_{k_1 \ldots \,k_s}
\,x_1^{k_1}\cdots x_s^{k_s} \;,
\label{lnxi}\ee
where $s$ is the number of species,
one can find the coefficients $f_{k_1 \ldots \,k_s}$ one by one.
We infer a general formula
based on explicit calculations for low orders,
\begin{equation}
f_{k_1\ldots \,k_r{\underbrace{\scriptstyle 0\ldots 0}_{s-r}}} =
(-1)^{r-1}
\frac{\prod_{j=1}^r \prod_{l=1}^{k_j-1}
\left( 1 - \frac{\sum_{n=1}^r \textstyle\mathstrut k_n g_{jn}}
{\textstyle\mathstrut l}\right)}
{\prod_{n=1}^r k_n} {\cal F}_{k_1\ldots \,k_r} \; ,
\label{f}
\end{equation}
where
\begin{equation}
{\cal F}_{k_1\ldots \,k_r} =
\sum_{{p_1,q_1,\ldots, \atop p_{r-1},q_{r-1}=1}}^r
\!\!\!\!\!\!\!\!\! {}^{\displaystyle *} \,\,\,\,\,
\left(\prod_{n=1}^{r-1} k_{q_{n}} g_{p_nq_n}\right) \; .
\label{F}
\end{equation}
Here all the numbers $k_1, \ldots , k_r$
are assumed to be different from zero, and
the $p_n$'s and $q_n$'s summed over in (\ref{F})
are constrained as follows:
(i) $p_1 < p_2 < \ldots < p_{r-1}$;
(ii) $p_n \neq q_n$;
(iii) it may not be $p_m = q_n$ and $q_m = p_n$;
(iv) at least one of the $q_n$'s has to be equal to the number
that is lacking in the set $\{p_1,\ldots,p_{r-1}\}$.

For example:
\be
{\cal F}_{k_1} & = & 1 \; ; \nonumber\\
{\cal F}_{k_1k_2} & = & k_1 g_{21} + k_2 g_{12} \; ; \nonumber\\
{\cal F}_{k_1k_2k_3} & = &
k_1^2g_{21}g_{31} + k_1k_2g_{21}g_{32} + k_3k_1g_{23}g_{31}
+ {\rm cycl.\; perm.}\;,
\label{Fs}\ee
where ``cycl.~perm.''~denotes terms obtained by
simultaneous cyclic permuta\-tions
of the subscripts of $g$'s and $k$'s. Hence one gets
\begin{equation}
f_{k0 \ldots 0}= \frac{1}{k}\prod_{l=1}^{k-1}
\left(1-\frac{kg_{11}}{l}\right)\;
\end{equation}
(cf.~\cite{I-IJMPA}) as well as
\be
f_{110\ldots 0}&=& -(g_{12}+g_{21}) \; ; \nonumber\\
f_{210\ldots 0}&=& -(g_{12}+2g_{21})(1-2g_{11}-g_{12})/2 \; ,
\nonumber\\
f_{1110\ldots 0}&=&
g_{21}g_{31} + g_{21}g_{32} + g_{23}g_{31} + {\rm cycl.~perm.}\;.
\ee
For two species and a symmetric matrix $g_{ab}$, the
coefficients (\ref{f}) reduce to those obtained in Ref.~\cite{SSS}.

One can also evaluate the coefficients of the expansions
\be
\ln \xi_a= \sum_{k_1 \ldots \,k_s}f^a_{k_1 \ldots \,k_s}
\,x_1^{k_1}\cdots x_s^{k_s} \;;
\label{lnxia}\ee
the answer is that
$f^a_{k_1\ldots \,k_s}$ is obtained from (\ref{f})
by replacing ${\cal F}_{k_1\ldots \,k_r}$ with
${\cal F}^a_{k_1\ldots \,k_r}$,
which in turn is obtained from (\ref{F}) by
restricting that $p_n \neq a$ for any $n$
(thereby fixing all the $p_n$'s). For example,
\be
{\cal F}^1_{k_1k_2} & = & k_1 g_{21} \; , \qquad
{\cal F}^2_{k_1k_2} = k_2 g_{12} \; ; \nonumber \\
{\cal F}^1_{k_1k_2k_3} & = &
k_1^2g_{21}g_{31} + k_1k_2g_{21}g_{32} + k_3k_1g_{23}g_{31} \;,
\quad {\rm etc.}
\label{F1}\ee

For a symmetric matrix $g_{ab}$, we observe an important property
\be
f^a_{k_1\ldots \,k_s}=\frac{k_a}{k_1+\cdots +k_s}
f_{k_1\ldots \,k_s}\;,
\label{f_a}\ee
which will be used in Sec.~\ref{SYMMAT} when deriving an equation of
state.

\section{Cluster and virial expansions}
\label{CLVIEXP}

The formulas of the previous section enable one to compute
the equation of state provided the single-particle spectrum
is known. Let the latter be
$\e_a^{(i)}=\e_a^{(0)}+\zeta^{(i)}$, where the $\zeta^{(i)}$'s
are common for all the species.
By summing (\ref{lnxi}) over single-particle
states ($\sum_i \ln \xi^{(i)} = \ln \Xi = -\beta\Omega$),
one then obtains a cluster expansion
\be
-\beta\Omega=
\sum_{k_1 \ldots \,k_s}b_{k_1 \ldots \,k_s}
\,z_1^{k_1}\cdots z_s^{k_s} \;, \quad z_a=e^{\beta\mu_a} \;
\label{clusterexpansion}\ee
with the cluster coefficients
\be
b_{k_1 \ldots \,k_s}=
Z'_1\bigl(K\beta \bigr)
e^{-\beta\sum_a k_a \e_a^{(0)}}
f_{k_1 \ldots \,k_s}\;,
\label{b_k}\ee
where $K = \sum_{n=1}^s k_n$ and
\be
Z'_1(\beta) = \sum_i e^{-\beta \zeta^{(i)}}
\label{Z1prime}\ee
is the (species-independent) ``shifted'' single-particle
partition function, corresponding to counting
the single-particle energy from the ground state.
One has $Z_1^a(\beta) \equiv \sum_i e^{-\beta\e_a^{(i)}} =
e^{-\beta \e_a^{(0)}}Z'_1(\beta)$.
Assume that $Z'_1(\beta)$ scales with the inverse tempera\-ture
$\beta$ as (cf.~\cite{IsO})
\be
Z'_1(K\beta)=\frac{Z'_1(\beta)}{K^{1+\delta}} \:.
\label{scaling}\ee
We will see below that
this scaling is relevant in the thermodynamic limit for various
physical systems.
The cluster coefficients (\ref{b_k}) then become
\be
b_{k_1 \ldots \,k_s}
=\frac{Z'_1(\beta)}{K^{1+\delta}}
e^{-\beta \sum_a k_a \e_a^{(0)}} f_{k_1 \ldots \,k_s} \;.
\label{b_k=}\ee
The expression for particle numbers
$N_a=z_a \frac{\D}{\D z_a}\ln \Xi$
reads
\be
N_a=\sum_{k_1 \ldots \,k_s} k_a \,b_{k_1 \ldots \,k_s}
\,z_1^{k_1}\cdots z_s^{k_s} \;.
\label{N_a}\ee
Using (\ref{f}) to determine the cluster coefficients (\ref{b_k=}),
one then deduces from (\ref{clusterexpansion}) and (\ref{N_a})
the ``virial expansion''
\be
-\beta\Omega=
\sum_{k_1 \ldots \,k_s}
 A_{k_1 \ldots \,k_s}
{{N_1^{k_1}\cdots N_s^{k_s}}\over
{[Z'_1(\beta)]^{k_1+\cdots+k_s-1}}}\;,
\label{`vir-exp'}\ee
where the dimensionless ``virial coefficients''
$A_{k_1 \ldots \,k_s}$ can {\it all} be evaluated
exactly\footnote{
Expressions for virial coefficients in terms of cluster coefficients
for two species may be found in Ref.~\cite{SSS}. We give here the
simplest formula of that kind for three species:
$A_{111} = 2 \tilde b_{011} \tilde b_{101} +
 2 \tilde b_{011} \tilde b_{110} +
 2 \tilde b_{101} \tilde b_{110} - 2 \tilde b_{111}$, where
$\tilde{b}_{k_1k_2k_3} = b_{k_1k_2k_3}/
b_{100}^{k_1}b_{010}^{k_2}b_{001}^{k_3}$,
as well as the associated formula
relating the cluster coefficient
to many-particle partition functions:
$b_{111} =
2 Z_{001} Z_{010} Z_{100} - Z_{100} Z_{011} -
Z_{010} Z_{101} - Z_{001} Z_{110} + Z_{111}$.
},
even if there is apparently no general expression for them.
Up to the third order, they are:
\be
A_{10\ldots0} &=& 1 \;;\nonumber\\
A_{20\ldots0} &=& -{2^{-2-\delta}}(1-2g_{11})\;,
\nonumber\\
A_{110\ldots0} &=&
2^{-1-\delta}}{(g_{12} + g_{21})\;;\nonumber\\
A_{30\ldots0} &=& (4^{-1-\delta }-2\cdot3^{-2-\delta })
-(4^{-\delta }-3^{-\delta })g_{11}(1-g_{11})\;,\nonumber\\
A_{210\ldots0} &=&
-{4^{-1-\delta }}{\left( g_{12} + g_{21} \right)
       \left( 2 - 4g_{11} - g_{12} - g_{21} \right) }
\nonumber\\&& {}+
 {3^{-1-\delta }}\left( g_{12} + 2g_{21} \right)
      {\left( 1 - 2g_{11} - g_{12} \right)} \;,\nonumber\\
A_{1110\ldots0} &=&
-2\cdot{3^{-1-\delta }}(g_{21}g_{31} + g_{21}g_{32} + g_{23}g_{31})
\nonumber\\&& {}
+ {4^{-1/2-\delta }}(g_{21}g_{31}+g_{21}g_{32}+g_{23}g_{31}+
g_{12}g_{13})
\nonumber\\&& {}+ {\rm cycl.~perm.}\;.
\label{A}\ee
Note that the mixed virial coefficient of order two vanishes for an
antisymmetric statistics matrix.

For a symmetric statistics matrix, the last two expressions become
\be
A_{210\ldots0}&=&-(4^{-\delta}-3^{-\delta
})g_{12}(1-2g_{11}-g_{12})\;,
\nonumber\\
A_{1110\ldots0}&=& 2(4^{-\delta }-3^{-\delta })
(g_{12}g_{13} + g_{12}g_{23} + g_{13}g_{23})\;.
\label{A-symm}
\ee
If in addition $\delta=0$, then
the virial coefficients of total order three do not depend on the
statistics parameters.
This case will be discussed in detail in the next section.

\section{Thermodynamics for a symmetric statistics matrix at
a constant density of states}
\label{SYMMAT}

The case of $g_{ab} = g_{ba}$ and $\delta =0$,
in fact, can be completely analyzed.
It is also of particular interest since it
means a $1/\beta$ scaling for $Z'_1(\beta)$,
i.e., a constant density of states in energy.
The particle numbers (\ref{N_a}) read
\be
N_a= Z'_1(\beta)\sum_{k_1 \ldots \,k_s}\frac{k_a}{k_1+\cdots +k_s}\,
f_{k_1 \ldots \,k_s}\,z_1^{k_1}\cdots z_s^{k_s} \;.
\label{Na}\ee
The statistics matrix being symmetric,
using Eq.~(\ref{f_a}) yields an important relation
\be
N_a=Z'_1({\beta})\ln \xi_a |_{x_b\to z_b\;{\rm for\: all}\;b}\;,
\label{N-xi}\ee
which enables one to use all the formulas from the previous section
provided the arguments of the functions $\xi_a$ and $w_a$,
the Gibbs factors $\{x_b\}$, are replaced by the fugacities $\{z_b\}$.

To deduce the equation of state, consider the derivatives
\be
-\frac{\D\,\beta\Omega}{\D N_a}=
\sum_b \frac{\D z_b}{\D N_a}\frac{N_b}{z_b} \;.
\label{D/D}\ee
{}From (\ref{N-xi}) we find
\be
\frac{\D z_b}{\D N_a}=\frac{(M^{-1})_{ba}}{Z'_1(\beta)}\;,
\label{dz/dN}\ee
where the matrix
$M_{ab}\equiv \frac{\D}{\D z_b} \ln \xi_a $
can be expressed, with the use of (\ref{xi-w}) and (\ref{dw/dx}), as
\be
M_{ab}=\frac{\xi_a}{z_b}(D^{-1})_{ab} \;,
\label{M=}\ee
and consequently
\be
(M^{-1})_{ba}=\frac{z_b}{\xi_a}D_{ba} \;.
\label{M^-1}\ee
Inserting this into (\ref{dz/dN}) and then the result into
(\ref{D/D}),
we obtain
\be
-\frac{\D\,\beta\Omega}{\D N_a}=\frac{N_a}{Z'_1(\beta)}
\frac{1}{e^{{N_a}/{Z'_1(\beta)}}-1} +\sum_b g_{ab}
\frac{N_b}{Z'_1(\beta)} \;,
\label{dOmega/dNa}\ee
where the symmetry of the statistics matrix was used.
Integrating (\ref{dOmega/dNa}) yields finally the equation of state
\be
-\Omega=\frac{1}{\beta Z'_1(\beta)}
\left\{
\sum_a \int_0^{N_a}\frac{\zeta_a \,
d\zeta_a}{e^{{\zeta_a}/{Z'_1(\beta)}}-1}+
\frac12 \sum_{ab} g_{ab}N_a N_b
\right \} \;.
\label{EqSt}\ee

Due to the scaling $Z'_1(\beta)\propto \beta^{-1}$, only
zero-temperature
terms in the last expression depend on the statistics parameters.
Equation (\ref{EqSt}) can be expanded to give both low-temperature
and
low-density (high-temperature) series.  In the former case,
restricting to the perturbative (behaving as powers of the
temperature) terms,
one obtains
\begin{equation}
-\Omega = E = \frac{\pi^2}{6} s { Z_0} T^2
+\frac{1}{2Z_0} \sum_{ab} g_{ab}N_a N_b \;,
\quad  Z_0\equiv {\beta Z'_1(\beta)} \;.
\label{E-lowT}\end{equation}
This in particular implies that the specific heat is
$C=\frac13 {\pi^2} s { Z_0} T $
to all orders in $T$, which is the same as
that for a mixture of  Fermi gases
(one should impose the condition $g>0$ to do the low-temperature
expansion).

In the opposite limit,
for small ${N_a}/{Z'_1(\beta)}$, one can expand (\ref{EqSt}) to
obtain
\begin{equation}
-\beta\Omega=\sum_a  N_a
+\frac12 \sum_{ab}(g_{ab}-\frac12 \delta_{ab})
\frac{N_a N_b}{Z'_1(\beta)}
+\sum_a \sum_{k=2}^{\infty}\frac{{\cal B}_k}{(k+1)!}
\left(\frac{N_a}{Z'_1(\beta)} \right)^{k+1} ,
\label{virexp}
\end{equation}
where
${\cal B}_k$ are the Bernoulli numbers  (${\cal B}_2=\frac16$,
${\cal B}_4=-\frac{1}{30}$, etc., ${\cal B}_{2l+1}=0$).
Thus, just as Eq.~(\ref{A}) shows, it is only the second order
virial coefficients that depend on the statistics matrix.
Eqs.~(\ref{EqSt})--(\ref{virexp}) are generalizations of the results
of Refs.~\cite{MS94,BhSen,IAMP,IsO} to the multispecies case.

\section{Examples}
\label{EXAMP}

In this section we consider particular
dispersion laws of particles,
mostly discussing examples which are related to integrable models.
For a single species, it has been shown that integrable models
with inverse square interactions, the
Calogero \cite{Calo} and Sutherland \cite{Sutherland} models,
are equivalent thermodynamically to systems of noninteracting
particles obeying exclusion statistics.
This naturally suggests that there should exist
generalizations of those models corresponding
to multispecies exclusion statistics.
It is in this sense that we refer below to the ``multispecies
Calogero and Sutherland models'' (which have yet to be discovered).
The equations of state found below for
noninteracting particles obeying exclusion statistics apply at
the same time to the conjectured multispecies integrable models,
thus providing guidelines for their search.

\subsection{Particles in a box}
\label{box}

Consider a gas of free particles in a $D$-dimensional box
of volume $V$.
In the quasicontinuous spectrum approximation,
when the momentum sum\-mation can be replaced by integration,
for particles with the dispersion law
$\e_a(p)=\e_a^{(0)}+\Lambda p^{\sigma}$ ($\sigma\neq 0$)
the shifted single-particle partition function reads \cite{IAMP,IsO}
\be
Z'_1(\beta)=
\frac{\Gamma(1+D/\sigma)V}
{ (2\sqrt{\pi}{})^D \Gamma(1+D/2)
(\Lambda\beta)^{D/\sigma} }\; ,
\label{Z1'}\ee
satisfying (\ref{scaling}) with $\delta = D/\sigma - 1$.
One has $\Omega=-PV$, where $P$ is the pressure,
and the expansion (\ref{`vir-exp'}) becomes the usual
virial expansion for a system in a box:
\be
\beta P = \sum_{k_1 \ldots \,k_s} a_{k_1 \ldots \,k_s}\,
\rho_1^{k_1}\cdots \rho_s^{k_s}, \quad  \rho_a=N_a/V \;,
\label{vir-exp-usual}\ee
where the (dimensional) virial coefficients are
\be
a_{k_1 \ldots \,k_s}=A_{k_1 \ldots \,k_s}
\left( {{V}\over {Z'_1(\beta)}} \right)^{k_1+\cdots +k_s-1}\:,
\label{a-dim}\ee
with $A_{k_1 \ldots \,k_s}$ defined in Sec.~\ref{CLVIEXP}.

\subsubsection{``Multispecies  Sutherland model''
($D=1$, $\sigma = 2$, $\delta = -1/2$)}
\label{Sutherland}

For a single species of particles with a quadratic dispersion law
$\e_a(p)=p^2/2m$ in one dimension, it is the Sutherland model
that realizes exclusion statistics.
For a conjectured multispecies version of that model,
the dimensional factor in (\ref{a-dim}) relating
the dimensional virial coefficients to the dimensionless ones is
$V/Z'_1(\beta)= \lambda_T$, where
$\lambda_T =\sqrt{2\pi \beta/m}$ is the thermal wavelength.
The single-species expressions agree with those obtained in
Refs.~\cite{dVO95com,IAMP}.

\subsubsection{Chiral fields on a circle
($D=1$, $\sigma = 1$, $\delta = 0$)}

The model of chiral fields on a circle \cite{IV96}
(a generalization of the model \cite{HLV} to the multispecies case)
is a field theoretical realization of exclusion statistics, with
a symmetric statistics matrix, for particles with linear dispersion
(above a gap) propagating in only one direction:
$\e_a(p)=\e^{(0)}_a+vp$, $p\geq 0$.
Due to the latter condition,
$Z'_1(\beta)=V/2\pi v \beta$, half of the value given by
(\ref{Z1'}).
Since $\delta=0$, the equation of state is given by (\ref{EqSt});
the low-temperature expansion (\ref{E-lowT})
agrees with that found in Ref.~\cite{IV96}.

\subsubsection{Nonrelativistic particles in 2 dimensions
($D=2$, $\sigma = 2$, $\delta = 0$)}
\label{2D-box}

Particles with a quadratic dispersion law in two dimensions
and a symmetric statistics matrix
(a conjectured model of Ref.~\cite{M96}, see Sec.~\ref{CONCLU})
also fall into the category
considered in Sec.~\ref{SYMMAT} so that the equation of state is
again (\ref{EqSt}), with $V/Z'_1(\beta)= \lambda_T^2$.

\subsection{Particles in a harmonic potential}
\label{harmwell}

We now consider systems of noninteracting nonrelativistic
particles in a harmonic well of strength $\omega$ in $D$ dimensions.
The single-particle partition function is
\be
Z_1^a(\beta)=
\frac{e^{-\frac12 D \beta\omega}}{(1-e^{-\beta\omega})^D}\,.
\label{Z1} \ee
Expanding this to the leading order in $\beta\omega$ results in
\be
Z'_1(\beta)\simeq
\frac{1}{(\beta\omega)^D}\;,
\label{Z1=}\ee
thus respecting the scaling (\ref{scaling}) with $\delta=D-1$
(with $D=1$ as the condition for the density of states to be
constant). This leads to the equation of state (\ref{`vir-exp'})
[and to (\ref{EqSt}) for $D=1$ and a symmetric statistics matrix].

If one considers the correction terms, of order
${\cal O}(1)$ in (\ref{Z1=}), they might lead to corrections
to the virial coefficients of very high order, $N$ and above.
However, if the virial expansion converges,
these corrections are negligible inside the radius of convergence.

Note that for particles in a harmonic potential,
there is no well defined volume occupied by the gas;
therefore, the equation of state is understood
as an equation relating the thermodynamic potential $\Omega$
to the particle numbers, temperature, and the harmonic
frequency $\omega$. Note also that the single-particle
dispersion law is linear ($\sigma=1$) here, because the
single-particle energy depends linearly on the quantum numbers.

\subsubsection{``Multispecies Calogero model''
($D=1$, $\sigma = 1$, $\delta = 0$)}
\label{Calogero}

For a single species with a linear dispersion law, the
Calogero model realizes exclusion statistics.
The ``multispecies Calogero model'', which would be
equivalent to a system of noninteracting
particles with a symmetric statistics matrix
in a one-dimensional harmonic potential \cite{M96},
would again fall into the class considered in Sec.~\ref{SYMMAT}
and be governed by the equation of state
(\ref{EqSt}), with $Z'_1(\beta) = 1/\beta\omega$.

\subsubsection{Harmonic potential in 2 dimensions
($D=2$, $\sigma = 1$, $\delta = 1$)}
\label{2D-harm}

For the same model as in \ref{2D-box} but involving a
harmonic well, i.e., a linear dispersion law,
whence $Z'_1(\beta) = 1/(\beta\omega)^2$,
the equation of state is the generic Eq.~(\ref{`vir-exp'})
with the ``virial coefficients'' (\ref{A}).

\medskip
Another important example, to be considered in detail
in the next section, is multispecies anyons in the LLL,
both in a harmonic potential and in a box.

\section{Multispecies anyons in the lowest Landau level}
\label{LLL}

The problem of anyons in the LLL,
being of relevance, in particular, to the FQHE,
was originally solved for a single species in Ref.~\cite{dVO94}.
The multispecies version of the
problem was addressed in Ref.~\cite{SSS}, with the restriction
that all the species have the same electric charge and mass, and the
equation of state in a box was derived. Here we revisit
this problem, allowing anyons  of different species to have
different magnitudes (but the same sign) of charges
and different masses. We derive
the equations of state in a box and in a harmonic potential,
showing how they fit into our general
framework, and provide a simple semiclassical picture.

\subsection{Spectrum in a harmonic potential}
\label{SPECTRUM}

The single-particle Hamiltonian in
a magnetic field and a harmonic potential is
\begin{equation}
H = -\frac{2}{m} \partial \bar{\partial }
- \omega_{{\rm c}} (z\partial - \bar{z}\bar{\partial })
+ \frac{m\omega_{{\rm t}}^{2}}{2} z\bar{z} \;
\label{Hsingle}
\end{equation}
($z=x+{\rm i}y$, $\partial = \partial /\partial z$,
$\bar{\partial } = \partial /\partial \bar{z}$),
and its spectrum is
\begin{equation}
E_{\ell n} = \left( \ell + \frac 12 \right)(\omega_{\rm t} -
\omega_{\rm c})
  + \left( n + \frac 12 \right)(\omega_{\rm t} + \omega_{\rm c}) \; ,
\label{Esingle}
\end{equation}
where $\omega_{{\rm c}} = -eB/2m$ ($eB<0$),
$\omega_{{\rm t}} = \sqrt{\omega_{{\rm c}}^2 + \omega^{2}}$,
with $\ell,n=0,1,2,\ldots$. The quantum numbers $\ell$ and $n$
are, respectively, the angular momentum and the number of
the Landau level. The LLL restriction, $n=0$, leads to
\begin{equation}
E_\ell = \omega_{\rm t} + \ell\varpi \;,
\label{ELLL}
\end{equation}
with $\varpi = \omega_{\rm t} - \omega_{\rm c}$.

In the multispecies problem, the charges $e_a$ and the
masses $m_a$ being different, so are the cyclotron frequencies
$\omega_{{\rm c}a}$. Let the harmonic
frequencies $\omega_{a}$ be different as well.
The many-body Hamiltonian becomes
\begin{equation}
H_N = \sum_{aj} \left[- \frac{2}{m_a} \partial_{aj}
\bar{\partial}_{aj}
- \omega_{{\rm c}a} (z_{aj}\partial_{aj} -
\bar{z}_{aj}\bar{\partial}_{aj})
+ \frac{m_a\omega_{{\rm t}a}^{2}}{2} z_{aj}\bar{z}_{aj} \right] \;
\label{Hmany}
\end{equation}
($j$ numbers particles of a given species,
$\partial_{aj} = \partial /\partial z_{aj}$).
Making an ansatz
\begin{equation}
\Psi_N = \tilde{\Psi}_N
\exp\left( -\sum_{aj} \frac{m_a\omega_{{\rm t}a}}{2}
z_{aj}\bar{z}_{aj}\right)
\label{psitilde}
\end{equation}
yields the Hamiltonian acting on $\tilde{\Psi}_N$,
\begin{equation}
\tilde{H}_N =
\sum_{aj} \left[- \frac{2}{m_a} \partial_{aj} \bar{\partial}_{aj}
+ (\omega_{{\rm t}a} - \omega_{{\rm c}a}) z_{aj}\partial_{aj}
+ (\omega_{{\rm t}a} + \omega_{{\rm c}a})
  \bar{z}_{aj}\bar{\partial}_{aj}
+ \omega_{{\rm t}a} \right] \; .
\label{Htilde}
\end{equation}
We now choose the $\omega_{a}$'s such that
$\varpi \equiv \omega_{{\rm t}a} - \omega_{{\rm c}a}$ be the same
for all the species. For all the $\omega_{a}^2$'s thus defined
to be positive (and to tend to zero,
rendering the system free, when $\varpi\to0$),
all the $\omega_{{\rm c}a}$'s have to be positive, consequently
the signs of all the charges have to be the same, such
that $e_{a}B < 0$. Then the LLL eigenfunctions
of the Hamiltonian (\ref{Htilde}) satisfying
the anyonic interchange conditions \cite{SSS} have the form
\begin{equation}
\tilde{\Psi}_N = \prod_{(aj)<(bk)} (z_{aj} - z_{bk})^{\alpha_{ab}}
\prod_a \left\{ \prod_j z_{aj}^{\ell_{aj}} \right\}_{\rm sym} \; ,
\label{psieigen}
\end{equation}
where $(aj)<(bk)$ means: $a<b$ or ($a=b$ and $j<k$), so that each pair
is counted only once; and symmetrization is performed
over the coordinates of particles of the same species only.
The energy of the state (\ref{psieigen}) is
\begin{equation}
E_{\{\ell_{aj}\}} = \sum_a N_a \omega_{{\rm t}a} +
\left[ \sum_{aj} \ell_{aj} + \sum_a \frac{N_a(N_a-1)}{2}\alpha_{aa}
+ \frac{1}{2}\sum_{ab} N_a N_b \alpha_{ab} \right] \varpi \; .
\label{eeigen}
\end{equation}
The LLL spectrum is obtained by letting the $\ell_{aj}$'s run from
0 to $\infty$ with the restriction $\ell_{aj} \le \ell_{a,j+1}$.
The partition function is
\begin{eqnarray}
Z_{N_1\ldots \,N_s} & = &
\exp \left[ -\left( \sum_a \frac{N_a(N_a-1)}{2}\alpha_{aa}
+ \frac{1}{2}\sum_{ab} N_a N_b \alpha_{ab} \right) \beta \varpi\right]
\nonumber \\ && \times
\prod_a \frac{e^{-N_a\beta\omega_{{\rm t}a}}}
{\prod_{k=1}^{N_a} \left( 1 - e^{k\beta\varpi} \right)} \;.
\label{partfun}
\end{eqnarray}

\subsection{Equation of state}
\label{EQNOFSTATE}

Let us now derive the equations of state for multispecies anyons
in the LLL both in a (species-dependent, as explained above)
harmonic potential and in a box
(labeling the associated cluster and virial coefficients
with superscripts $\omega$ and $V$, respectively).

The cluster expansion for the system in a harmonic potential reads
\begin{equation}
\ln \Xi = \sum_{k_1 \ldots \,k_s} b_{k_1\ldots \,k_s}^{\omega}
z_1^{k_1}\cdots z_s^{k_s} \; ,
\label{lnXi}
\end{equation}
where the cluster coefficients $b_{k_1\ldots \,k_s}^{\omega}$
can be expressed in terms of the partition functions
in the standard way, by matching (\ref{lnXi}) to
$\Xi = \sum_{k_1 \ldots \,k_s} Z_{k_1\ldots\,k_s} z_1^{k_1} \cdots
z_s^{k_s}$, and the partition functions are given by
(\ref{partfun}).

We are interested in going to the thermodynamic limit.
In accordance with the aforesaid, the harmonic frequencies are
$\omega_a^2 = (\varpi + \omega_{{\rm c}a})^2 - \omega_{{\rm c}a}^2$,
where $\omega_{{\rm c}a} = -e_aB/2m_a$. The thermodynamic limit is
understood as $\varpi \to 0$, so that
$\omega_{a}^{2} \simeq 2\varpi\omega_{{\rm c}a} \to 0$.
To the leading order in $\beta \varpi$ and
$\omega_a/\omega_{{\rm c}a}$, the cluster coefficients are
\begin{equation}
b_{k_1\ldots \,k_s}^{\omega} = \frac{ e^{- \beta \sum_{a} k_a
\omega_{{\rm c}a}}}
{(k_1 + \cdots + k_s)\beta\varpi}
f_{k_1\ldots \,k_s}|_{g_{ab}\to\alpha_{ab}} \; .
\label{clus-omega}
\end{equation}
Appearance of the coefficients  $f_{k_1\ldots \,k_s}$ (\ref{f}) in this
expression indicates that exclusion statistics is present in
the system.
Noting that in the case at hand
$Z'_1(\beta)=1/\beta \varpi$, we see
that $Z'_1(\beta)$ and $b_{k_1\ldots \,k_s}^{\omega}$
satisfy (\ref{scaling}) and (\ref{b_k=}), respectively,
with $\e_a^{(0)}=\omega_{{\rm c}a}$ and $\delta =0$
(constant density of states).
We can thus conclude that the equation of state for LLL
anyons in the harmonic potential considered coincides
with (\ref{EqSt}), generalizing the result
of Ref.~\cite{IsO} for a single species.

Now we turn to the equation of state for the same system in a box.
There is a general procedure of deriving the equation of state
in a box starting from the cluster coefficients
in a harmonic potential (originally proposed for
anyons \cite{CGO89} and then put on general grounds \cite{O92}).
Here we follow the lines of Ref.~\cite{O92}.
Assuming that the external potential varies slowly in space,
the relation $\ln \Xi =\beta PV$ valid in a box is replaced by
\begin{equation}
\ln\Xi = \beta \int d^D r \, P({\bf r}) \; ,
\label{Xiinhom}
\end{equation}
where $P({\bf r})$ is the local pressure. The latter
is evaluated within a small volume where the external potential
$\Phi_a({\bf r})$ acting on the particles of the $a$-th species
is approximately constant, using the equation of state
for particles in a box with the chemical potential
$\mu_a$ replaced by $\mu_a - \Phi_a({\bf r})$:
\begin{equation}
\beta P({\bf r}) = \frac{1}{V} \sum_{k_1 \ldots\,k_s}
b_{k_1\ldots\,k_s}^{V}
\left( z_1 e^{-\beta \Phi_1({\bf r})}\right)^{k_1} \cdots
\left( z_s e^{-\beta \Phi_s({\bf r})}\right)^{k_s} \; .
\label{pofr}
\end{equation}
Substituting this into (\ref{Xiinhom}) and comparing the
result to (\ref{lnXi}), one gets
\begin{equation}
b_{k_1\ldots\,k_s}^V = V \frac{b_{k_1\ldots\,k_s}^{\omega}}
{\int d^D r \, e^{-\beta \sum_a k_a \Phi_a({\bf r})}} \; .
\label{bl1ls}
\end{equation}
For $D=2$ and $\Phi_a({\bf r}) = m_a\omega_a{\bf r}^2/2$,
this turns into $b_{k_1\ldots\,k_s}^V =
\beta V (k_1 m_1 \omega_1^2 + \cdots + k_s m_s \omega_s^2)
b_{k_1\ldots\,k_s}^{\omega}/2\pi$.
In going to the thermodynamic limit, one has
$m_a \omega_a^2 \simeq 2m_a \varpi \omega_{{\rm c}a} = 2\pi \varpi
\rho_{{\rm L}a}$,
where $\rho_{{\rm L}a} = m_a \omega_{{\rm c}a}/\pi $ is the
Landau level degeneracy per unit area.
Hence, we find the relation between
the cluster coefficients of LLL anyons in a box and in a
harmonic potential (the ``thermodynamic limit prescription''):
\begin{equation}
b_{k_1\ldots\,k_s}^V = V(k_1 \rho_{{\rm L}1} + \cdots +
k_s \rho_{{\rm L}s})
\beta \varpi b_{k_1\ldots\,k_s}^{\omega} \; .
\label{clusLLL}
\end{equation}

Substituting (\ref{clus-omega}), one finds
\begin{equation}
b_{k_1\ldots\,k_s}^V = V\frac{k_1 \rho_{{\rm L}1}
+ \cdots + k_s \rho_{{\rm L}s}}
{k_1 + \cdots + k_s} e^{- \beta \sum_{a} k_a \omega_{{\rm c}a}}
f_{k_1\ldots \,k_s}|_{g_{ab}\to \alpha_{ab} } \; ;
\label{clusfin}
\end{equation}
hence, the virial coefficients as in (\ref{vir-exp-usual}) are
\begin{equation}
a_{k_1\ldots\,k_s} = -\frac{(k_1 + \cdots + k_s - 1)!}{k_1!
\cdots k_s!}
\sum_a \frac{[(\frac{\alpha_{aa}-1}{\alpha_{aa}})^{k_a}-1]
\prod_{b} \alpha_{ab}^{k_b}}{\rho_{{\rm L}a}^{k_1+\cdots+k_s-1}} \; ,
\label{virfin}
\end{equation}
which implies the following equation of state
for LLL anyons in a box:
\begin{equation}
\beta P = \sum_{a} \rho_{{\rm L}a}
\ln \left( 1 + \frac{\rho_a/\rho_{{\rm L}a}}
{1 - \sum_b \alpha_{ab} \rho_b/\rho_{{\rm L}a}}\right) \; .
\label{eqst-LLL-box}
\end{equation}

This equation also shows that the statistical
mechanics of the above system of anyons is
governed by exclusion statistics. Indeed, starting directly from
Haldane's definition (\ref{StatWeight})
and assuming that all particles of the same
species have the same energy $\e_a$ ($\omega_{{\rm c}a}$
in the case at hand),
one easily derives (see also Ref.~\cite{Wu94})
the equation of state which coincides with  (\ref{eqst-LLL-box})
upon changes $G_a/V\to \rho_{{\rm L}a}$ and $g_{ab} \to \alpha_{ab}$.
[For a species-independent $\rho_{{\rm L}a} \equiv \rho_{\rm L}$,
Eq.~(\ref{clusfin}) is of course again a special case
of (\ref{b_k=}), with $\e_{a}^{(0)} = \omega_{{\rm c}a}$,
$Z'_1(\beta)=\rho_{\rm L}V$, $\delta=-1$.]

It is possible to give a mean-field interpretation \cite{dVO95,SSS}
of Eq.~(\ref{eqst-LLL-box}) in the spirit of the Chern-Simons
model. In that model, the statistics parameters are given by
\begin{equation}
\alpha_{ab} = \frac{e_a \phi_b}{2\pi}\;,
\label{alphaab}
\end{equation}
where $\phi_{b}$ is the Chern-Simons flux, whose proportionality
to the charge $e_b$ ensures that $\alpha_{ab}$ is symmetric%
\footnote{Note that if the flux comes from a real solenoid,
the corresponding formula is
$\alpha_{ab} = (e_a \phi_b + e_b \phi_a)/2\pi$,
differing essentially by a factor of two from the above \cite{GMW}.
We thank Alfred Goldhaber for elucidating the point.}.
Smearing out the fluxes and adding them to the external
magnetic field changes the Landau density $\rho_{{\rm L}a}$
into $\rho_{{\rm L}a}^{\rm eff} = \rho_{{\rm L}a} -
\sum_b \alpha_{ab}\rho_{b}$. Now, the quantity under the
logarithm in (\ref{eqst-LLL-box}) is
$1 + \rho_a/\rho_{{\rm L}a}^{\rm eff}$,
which is what one gets by setting all $\alpha_{ab}$'s
to zero and replacing $\rho_{{\rm L}a}$ with
$\rho_{{\rm L}a}^{\rm eff}$.
Equation (\ref{eqst-LLL-box}) is different from the one conjectured
previously in Ref.~\cite{dVO95}, although they coincide for
a species-independent $\rho_{\rm L}$.

\subsection{Semiclassical quantization}
\label{SEMICL}

Further insight into the nature of exclusion statistics in the
system at hand can be gained by considering it semiclassically.
The classical equation of motion of a particle in a magnetic
field and a harmonic potential is, in our notation,
\begin{equation}
\ddot{\bf r} = -\omega ^2{\bf r}
- 2\omega_{\rm c} \dot{\bf r} \times {\bf e}_z \; .
\label{classeq}
\end{equation}
Normal modes, obviously, correspond to circles,
with $\dot{\bf r} \times {\bf e}_z = -\Omega {\bf r}$
($eB<0$, so we have chosen positive $\Omega$ to
correspond to clockwise rotation)
and $\ddot{\bf r} = -\Omega^2 {\bf r}$. Substituting
into the above brings up two solutions,
\begin{equation}
\Omega = \pm \omega_{\rm t} - \omega_{\rm c} \; ,
\label{Omega}
\end{equation}
with $\omega_{\rm t}$ defined as above [cf.~(\ref{Esingle})].
The upper sign corresponds to the lifted degeneracy
of Landau levels, the lower one
to Landau excitations; opposite signs of $\Omega$ mean that the
directions of rotation corresponding to the two modes are opposite.
The LLL restriction means that only the former,
$\Omega = \varpi$ mode is excited,
so that single-particle orbits are circles, the angular
momentum is quantized in integer units, and the energy is
\begin{equation}
E_\ell = \ell\varpi \; .
\label{eclass}
\end{equation}
For a many-body system, the orbits are
concentric circles and can be ordered by increasing value of
angular momentum (or radius). Putting a flux $\phi$ inside the orbit
of a charge $e$ increases the allowed values of the kinetic angular
momentum
by $\alpha = e\phi/2\pi$. Consequently, the quantization condition now
is that the angular momentum $\tilde{\ell}_{aj}$ of a particle
is an integer plus additions from all particles whose
orbits are inside (cf.~\cite{RR92}):
\begin{equation}
\tilde{\ell}_{aj} = \ell_{aj} +
\sum_{bk} \alpha_{ab} \theta (\tilde{\ell}_{aj} -
\tilde{\ell}_{bk}) \;
\label{ltilde}
\end{equation}
[Eq.~(\ref{alphaab}) has been taken into account],
where $\ell_{aj}$ are integers and $\theta$ is the step function
defined here as $\theta (x)=1$ for $x>0$,
$\theta (x)=0$ for $x<0$, and $\theta (0)=1/2$.
The total energy is $E = \sum_{aj} E_{\ell_{aj}}$, and substituting
(\ref{eclass}) and (\ref{ltilde}) herein yields the correct result
(\ref{eeigen}), save for zero oscillations.

Eqs.~(\ref{eclass})--(\ref{ltilde}) have recently appeared in
Ref.~\cite{IV96} as defining exclusion statistics (in a somewhat
different geometry); basically, they are the Bethe ansatz equations.

\section{Concluding remarks}
\label{CONCLU}

We have worked out in detail the statistical mechanics
and thermodynamics for multispecies exclusion statistics
starting from Haldane's combinatorial formula.
As of now, two main groups of models are known to
realize exclusion statistics: the two-dimensional
LLL anyon model and one-dimensional integrable models.
As for the first one, we have demonstrated that exchange
statistics is equivalent to exclusion statistics if
the charges of the particles are all of the same sign.
The case when charges of both signs are present,
which would be relevant, in particular,
to quasielectrons and quasiholes in the FQHE,
has yet to be investigated.
In fact, the spectrum (\ref{eeigen}) does not
lead to the thermodynamics considered for a nonsymmetric
matrix $g_{ab}$, and it is unknown whether
the LLL spectrum of anyons
with different signs of charges does so.

Concerning integrable models,
the most interesting question is as follows:
What would be the generalization
of the Calogero and Sutherland models that would realize
exclusion statistics in the multispecies case?
The virtual coincidence of the energy levels
of single-species LLL anyons and of the Calogero model
\cite{dVO95} suggests that the energy levels (\ref{eeigen})
for the multispecies case might at the same time be the
energy levels of some integrable model.
A similar argument applies to the Sutherland model,
for which the generalized momenta $\ell$ would still
satisfy an equation like (\ref{ltilde}), by the Bethe
ansatz argument, but now with a quadratic dependence
of the energy of free particles on those momenta.

One should note that
different microscopic spectra may result in the same thermodynamics.
In Ref.~\cite{dV-NPB}, a quantum mechanical spectrum for
a single-species two-dimensional system was proposed which
leads to an equation of state identical to that of two-dimensional
exclusion statistics particles with quadratic dispersion
(see Subsec.~\ref{2D-box}). The principal difference
of our approach from that of Ref.~\cite{dV-NPB} is
that the  spectrum (\ref{eeigen}) is derived from the rule
of the kind (\ref{ltilde}) for the generalized (``renormalized''
\cite{Kawakami}) momenta (a generalization of the rule (\ref{ltilde})
can also be given for a two-dimensional system \cite{M96}).
In terms of these momenta the energy has a
form corresponding to a free system [cf.~(\ref{eclass})], and the
effect of statistics thus amounts to changing  the bare momenta to
their ``renormalized'' values determined by equations of the type
(\ref{ltilde}), in accordance with the definition of multispecies
exclusion statistics by the Bethe ansatz-like equations
\cite{IV96}.
Along the above lines, the spectrum (\ref{eeigen})
and its two-dimensional generalization may be interpreted
as the spectra of the systems discussed in Subsecs.~\ref{Calogero}
and \ref{2D-box}, respectively \cite{M96}.

As was recently discussed, the generalization
of the Calogero-Sutherland mo\-del which keeps the interaction
two-body and only allows different couplings between particles of
different species  implies constraints on possible values of the
coupling parameters already
at the level of the ground state of the system \cite{SenNPB96}.
One might expect that the multispecies  generalizations
of the Calogero and Sutherland models which would reveal generic
mutual exclusion statistics should involve
three-body interactions, in the spirit of the approach of
Ref.~\cite{FO} which exploits analogies with anyons.

\section*{Acknowledgements}

We would like to thank Diptiman Sen for a discussion where he
suggested that a system of anyons considered
in Sec.~\ref{LLL} might be solvable.
We wish to thank the Senter for H\o yere Studier in Oslo,
where this work was initiated, for kind hospitality and
financial support.
In deriving analytic expressions for coefficients of the expansions,
{\it Mathematica} was used.

\end{document}